\RequirePackage{fix-cm}
\documentclass[smallextended]{svjour3}       
\smartqed  
\usepackage{graphicx}
\begin{document}

\title{Trade Network Reconstruction and Simulation with Changes in Trade Policy
}


\author{Yuichi Ikeda         
 \and
          Hiroshi Iyetomi
}


\institute{Y. Ikeda \at
              (To whom correspondence should be addressed) \\
              Graduate School of Advanced Integrated Studies in Human Survivability, Kyoto University \\
              1 Yoshida-Nakaadachi-cho, Sakyo-ku, Kyoto-shi, Kyoto 606-8306, JAPAN \\              
Tel.: +81-75-762-2102\\
              \email{ikeda.yuichi.2w@kyoto-u.ac.jp} \\
           \and
           H. Iyetomi \at
              Graduate School of Science and Technology, Niigata University \\  
              8050, Ikarashi 2-no-cho, Nishi-ku, Niigata 950-2181,JAPAN
}

\date{Received: date / Accepted: date}

\maketitle

\begin{abstract}

The interdependent nature of the global economy has become stronger with increases in international trade and investment. 
We propose a new model to reconstruct the international trade network and associated cost network by maximizing entropy based on local information about inward and outward trade. 
We show that the trade network can be successfully reconstructed using the proposed model. 
In addition to this reconstruction, we simulated structural changes in the international trade network caused by changing trade tariffs in the context of the government's trade policy.
The simulation for the FOOD category shows that import of FOOD  from the U.S. to Japan increase drastically by halving the import cost.  
Meanwhile, the simulation for the MACHINERY category shows that exports from Japan to the U.S. decrease drastically by doubling the export cost, while exports to the EU increased.

\end{abstract}

\section{Introduction}
\label{intro}

In this era of economic globalization, most national economies are linked by international trade and consequently form a complex global economic network. 
The interdependent nature of the global economy has become stronger with increases in international trade and investment. 
In Japan, it is expected that many small and medium enterprises will achieve higher economic growth in a free environment created by the establishment of economic partnership agreement (EPA), such as the Trans-Pacific Partnership (TPP). 

Apart from the economy, it is known that various collective motions exist in natural science.
The phenomenon of collective motion is caused by strong interactions between constituent elements (agents) in a system. 
Interesting collective motions should emerge in a global economy under trade liberalization.

This paper is organized as follows. In section \ref{PreStudy}, preceding studies are reviewed. In section \ref{Model}, we propose a new model to reconstruct the international trade network and associated cost network by maximizing entropy based on local information about inward and outward trade flows. Subsequently, in section \ref{Results}, we will show the results of trade network reconstruction using the proposed model. In addition to these reconstructions, we will show the results for the effect of reduced/increased trade tariffs and trade barriers on the change in the community structure of the international trade network. Section \ref{Conclusion} concludes the paper.

\section{Preceding Studies}
\label{PreStudy}

International trade has been studied from the perspective of network science. Various topological properties and temporal evolution were studied \cite{Garlaschelli2005} \cite{Bhattacharya2008} \cite{Fagiolo2009} \cite{Barigozzi2010} \cite{Barigozzi2011} \cite{Benedictis2011} \cite{Leeo2011} \cite{Deguchi2014} \cite{Ikeda2014} \cite{Ikeda2015} \cite{Ikeda2016} \cite{Aoyama2017}.
Early studies concentrated on the simple topological properties of weighted directed trade networks between countries. 
The network study gradually extended to involve community structures, dynamical properties including synchronization of the business cycle, commodity-specific analysis, and industry sector-specific analysis.
Recent studies discuss topics with a practical interest, such as the spreading of economic crises and network controllability.

The gravity model of international trade has been widely used in traditional international economics. Quantitative studies of the gravity model were conducted to establish a microscopic foundation of the model \cite{Bergstrand1985} \cite{Dueñas2013}.
The gravity model was also generalized to explain traffic network by considering traffic cost \cite{Wilson1967}. This model is applicable to various spatial flow problems, including transportation, human mobility, domestic and international trade.  

Based on the above pioneering studies of international trade in network science, reconstruction of a trade network is recognized as a promising new direction in this field \cite{Mastrandrea2014} \cite{Cimini2015a}.
Here, reconstruction means that construction of the whole network, i.e. the direction and weight of each link between nodes, from given local information about each node, i.e. the degree and strength of each node. The concept of reconstruction was applied to analyze systemic risk in financial networks and reconstruction of international trade flows by commodity and industry \cite{Cimini2015b} \cite{Ikeda2017}.

\section{Reconstruction Model}
\label{Model}

First, we develope a reconstruction model for the economic network and apply the model to the international trade network in which nodes and links are countries and amount of bilateral trade, respectively. 
We then describe an algorithm to reconstruct the associated cost network layer by estimating bilateral trade cost $c_{ij}$ which does not explicitly appear in the actual data.

\subsection{Ridge Entropy Maximization}
\label{RidgeEntropy}

Suppose that total exports $E_i$ and total imports $I_i$ are known for country $i$. 
First, we maximize configuration entropy $S$ under the given total exports and total imports.

\paragraph{Convex Optimization} 
\label{Convex}

Configuration entropy $S$ is written using bilateral trade $t_{ij}$ between country $i$ and $j$ as follows,
\begin{equation}
 S = log \frac{ \left( \sum_{ij} t_{ij} \right) ! }{ \prod_{ij} t_{ij} ! } \approx \left( \sum_{ij} t_{ij} \right) \log \left( \sum_{ij} t_{ij} \right) - \sum_{ij} \left( t_{ij} \log t_{ij} \right).
\label{entropy1}
\end{equation}
Here an approximation is applied to factorial $!$ using Stirling's formula.   
The first term of R.H.S. of Eq.(\ref{entropy1}) does not change the value of $S$ by changing $t_{ij}$ because $\sum_{ij} t_{ij}$ is constant. Consequently, we have a convex objective function:
\begin{equation}
 S = - \sum_{ij} \left( t_{ij} \log t_{ij} \right).
\label{entropy2}
\end{equation}
Entropy $S$ is to be maximized with the following constraints:
\begin{equation}
 E_i = \sum_j t_{ij},
\label{TotExp}
\end{equation}
\begin{equation}
 I_j = \sum_i t_{ij}.
\label{TotImp}
\end{equation}
\begin{equation}
 G = \sum_{ij} t_{ij}
\label{TotTrade}
\end{equation}
\begin{equation}
 t_{ij} \geq 0
\label{NonNegative}
\end{equation}
Here, constraints Eq. (\ref{TotExp}) and Eq. (\ref{TotImp}) correspond to local information about each node.

\paragraph{Sparse Modeling} 
\label{Sparse}

The accuracy of the reconstruction will be improved using the sparsity of the international trade network.
We note that we have two different types of sparsity here. 
The first is characterized by the skewness of the observed bilateral import and export distributions. Import and export distributions are known to be log-normal distributions \cite{Barigozzi2010}. 
The second type of sparsity is characterized by the skewness of the observed in-degree and out-degree distributions. This means that a limited fraction of nodes have a large number of links and most nodes have a small number of links and consequently the adjacency matrix of international trade is sparse.

To take into account the first type of sparsity,
the objective function (\ref{entropy2}) is modified by applying the concept of Lasso (least absolute shrinkage and selection operator) \cite{Tibshirani1996}  \cite{Breiman1995} \cite{Hastie2008} to our convex optimization problem. 

In the case of Lasso or L1 regularization, our problem is formulated as the maximization of objective function $z$: 
\begin{equation}
 z(t_{ij}) = S - \sum_{ij} \left| t_{ij} \right| = - \sum_{ij} \left( t_{ij} \log t_{ij} \right) - \beta' \sum_{ij} \left| t_{ij} \right|
\label{Lasso}
\end{equation}
with local constraints. Here the second term of R.H.S. of Eq. (\ref{Lasso}) is L1 regularization. $\beta'$ is a control parameter. However the L1 regularization term $\sum_{ij} \left| t_{ij} \right|$ is constant in our problem. Therefore, we need a variant of the Lasso concept, e.g. ridge or L2 regularization.

By considering this fact, our problem is reformulated as the maximization of objective function $z$:
\begin{equation}
 z(t_{ij}) = S - \sum_{ij} t_{ij}^2 = - \sum_{ij} \left( t_{ij} \log t_{ij} \right) - \beta' \sum_{ij} t_{ij}^2
\label{Ridge}
\end{equation}
with local constraints. Here the second term of R.H.S. of Eq. (\ref{Ridge}) is L2 regularization.

\paragraph{Correspondence Relationship to Thermodynamnics} 
\label{Thermodynamics}

In the theory of thermodynamics, the equilibrium of a system is obtained by minimizing thermodynamic potential $F$:
\begin{equation}
 F = E - TS
\label{FreeEnergy}
\end{equation}
where $E$, $T$, and $S$ are internal energy, temperature, and entropy, respectively. Eq. (\ref{FreeEnergy}) is rewritten as a maximization problem as follows,
\begin{equation}
 z \equiv -\frac{1}{T} F = S - \frac{1}{T} E.
\label{ObjtFcun}
\end{equation}
We note that Eq. (\ref{ObjtFcun}) has the same structure to Eq. (\ref{Ridge}). Thus we interpret the meaning of control parameter $\beta'$ and L2 regularization as inverse temperature and internal energy, respectively.

In summary, we have a ridge entropy maximization problem: 
\begin{equation}
\renewcommand{\arraystretch}{2.3}
\begin{array}{ll}
\mbox{maximize   } & z (p_{ij}) = - \sum_{ij} \left( p_{ij} \log p_{ij} \right) - \beta' \sum_{ij} p_{ij}^2 \\ 
\mbox{subject to   } & G = \sum_{ij} t_{ij} \\
                       & \frac{E_i}{G} = \sum_j \frac{t_{ij}}{G} = \sum_j p_{ij} \\ 
                       & \frac{I_j}{G} = \sum_i \frac{t_{ij}}{G} = \sum_i p_{ij} \\
                       & t_{ij} \geq 0 \\ 
\end{array}
\label{ConvexFormulation}
\end{equation}
The second type of sparsity, which is characterized as the sparsity of the adjacency matrix, will be discussed in section \ref{ResultReconstruction}.

\subsection{Reconstructing the Cost Network Layer}
\label{Cost}

%
\begin{figure*}
  \begin{center}
  \includegraphics[width=0.8\textwidth]{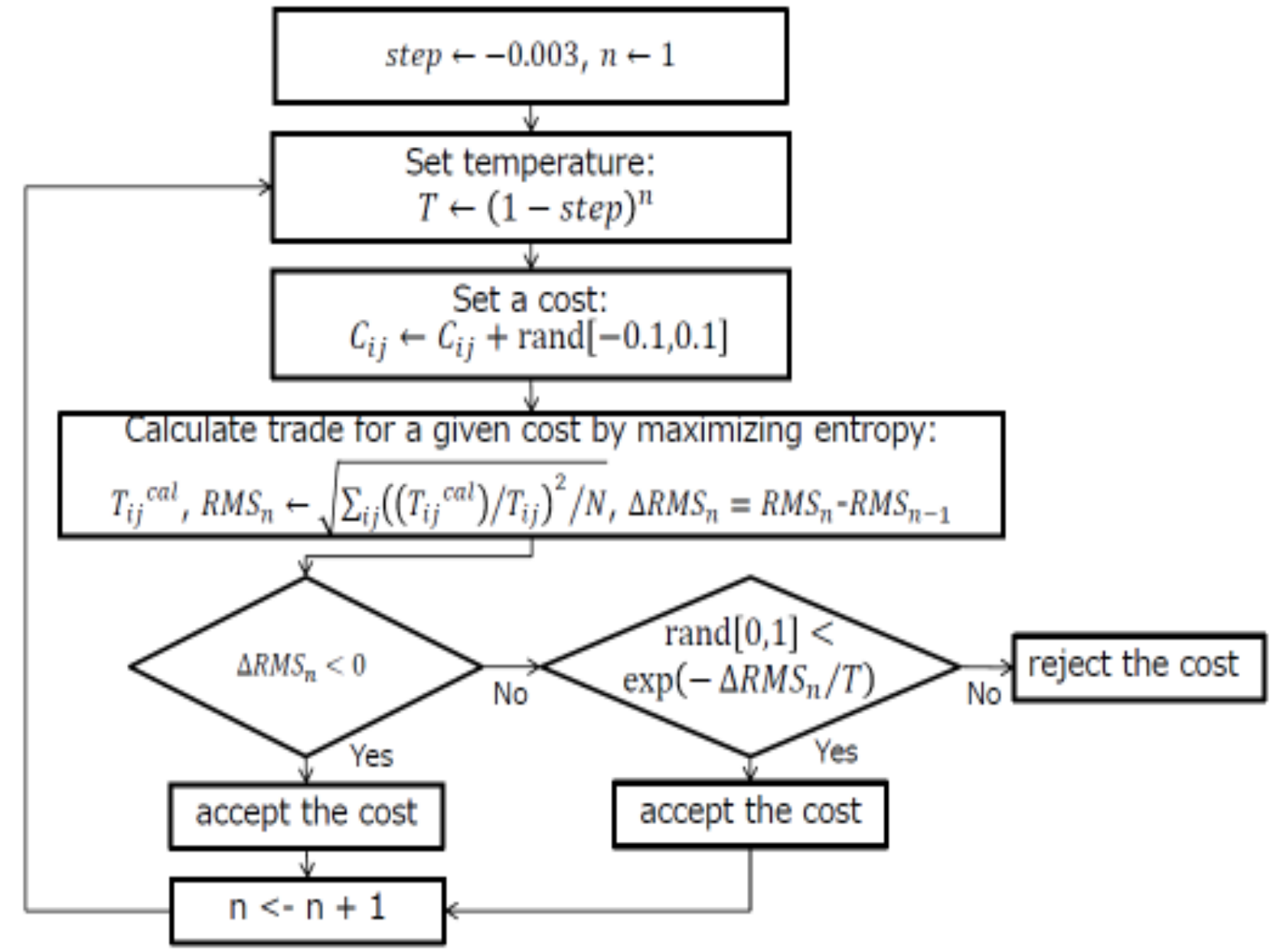}
\caption{The algorithm for reconstruction of the cost network layer using simulated annealing. The amount of trade is calculated using the iterative relationships in Eqs. (\ref{ReconstTrade}) - (\ref{CoeffB}) for the given trade cost. Trade cost is initially generated randomly at an initial high temperature and is converged to optimal values as the system is annealed. }
\label{Fig:Algorithm}
  \end{center}
\end{figure*}

In this subsection, we reconstruct the associated cost network layer by estimating bilateral trade cost $c_{ij}$, which does not appear explicitly in the actual data. Here, we note that trade cost $t_{ij}$ includes transportation cost, custom duty, and other non-tariff barriers. 
We formulate the problem to reproduce the reconstructed trade network, which is obtained by solving problem in Eq. (\ref{ConvexFormulation}), by taking into account trade cost $c_{ij}$.

Iterative relations for export $t_{ij}$ from country $i$ to $j$ are obtained by maximizing entropy $S$ in Eq. (\ref{entropy1}) with the constraints of Eq. (\ref{TotExp}) and Eq. (\ref{TotImp}) using the Lagrange multiplier method \cite{Wilson1967}:
\begin{equation}
  t_{ij} = A_i B_j E_i I_j \exp \left( - c_{ij} \right)
\label{ReconstTrade}
\end{equation}
\begin{equation}
  A_i = \left[ \sum_j B_j I_j \exp \left( - c_{ij} \right)  \right]^{-1}
\label{CoeffA}
\end{equation}
\begin{equation}
  B_j = \left[ \sum_i A_i E_i \exp \left( - c_{ij} \right) \right]^{-1}
\label{CoeffB}
\end{equation}
We note that bilateral trade cost $c_{ij}$ has to be given in advance to use these iterative relations. 

Bilateral trade cost $c_{ij}$ is estimated by combining iterative relations \cite{Wilson1967} in Eqs. (\ref{ReconstTrade}) - (\ref{CoeffB}) and simulated annealing \cite{Kirkpatrick1983} \cite{Kirkpatrick1984}.
The algorithm to estimate $c_{ij}$ is shown in Fig. \ref{Fig:Algorithm} \cite{Ikeda2017}.
The amount of trade is calculated using the iterative relations in Eqs. (\ref{ReconstTrade}) - (\ref{CoeffB}) for the given trade cost. Trade cost is initially generated randomly at an initial high temperature and is updated by adding a small change at each iteration step $n$. By repeating this procedure trade cost converges to optimal values as the system is annealed.

\section{Results and Discussions}
\label{Results}

\subsection{Reconstruction of the Trade Network}
\label{ResultReconstruction}

We reconstructed the international trade network by solving the ridge entropy maximization problem in Eq. (\ref{ConvexFormulation}). 
We used NBER-UN data which records trade amounts between bilateral countries for each type of commodities at year 2000 \cite{Feenstra2000}.

\begin{figure}[htbp]
 \begin{minipage}{0.5\hsize}
  \begin{center}
   \includegraphics[width=58mm]{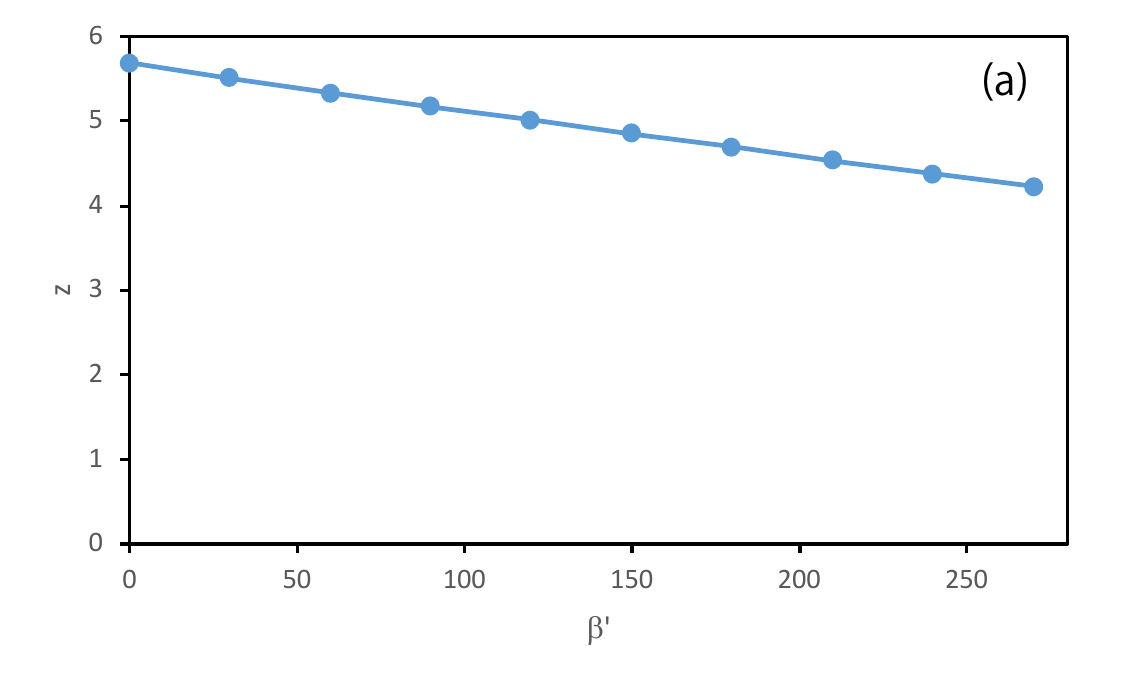}
  \end{center}
 \end{minipage}
 \begin{minipage}{0.5\hsize}
  \begin{center}
   \includegraphics[width=58mm]{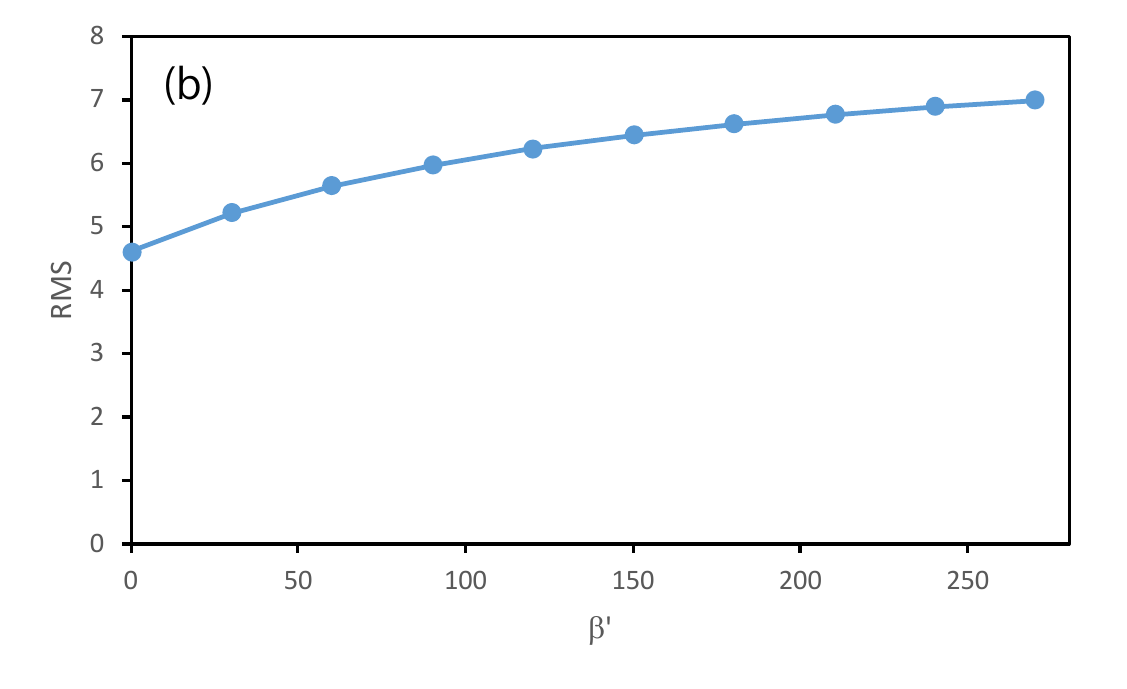}
  \end{center}
 \end{minipage} \\
 \begin{minipage}{0.5\hsize}
  \begin{center}
   \includegraphics[width=58mm]{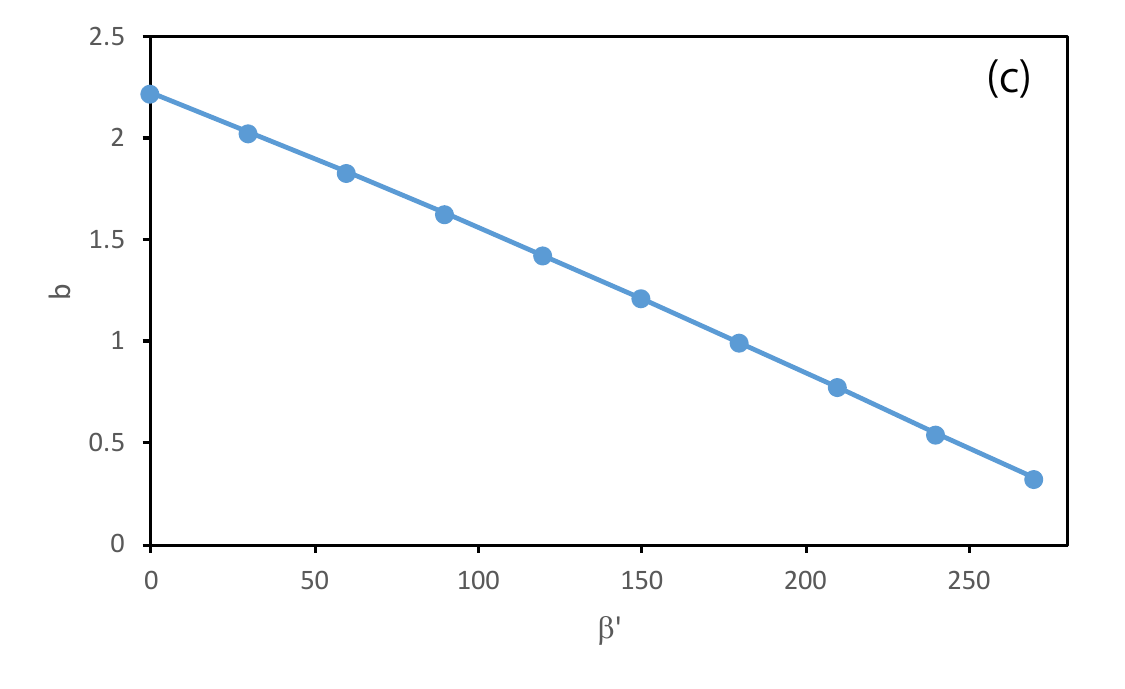}
  \end{center}
 \end{minipage}
 \begin{minipage}{0.5\hsize}
  \begin{center}
   \includegraphics[width=58mm]{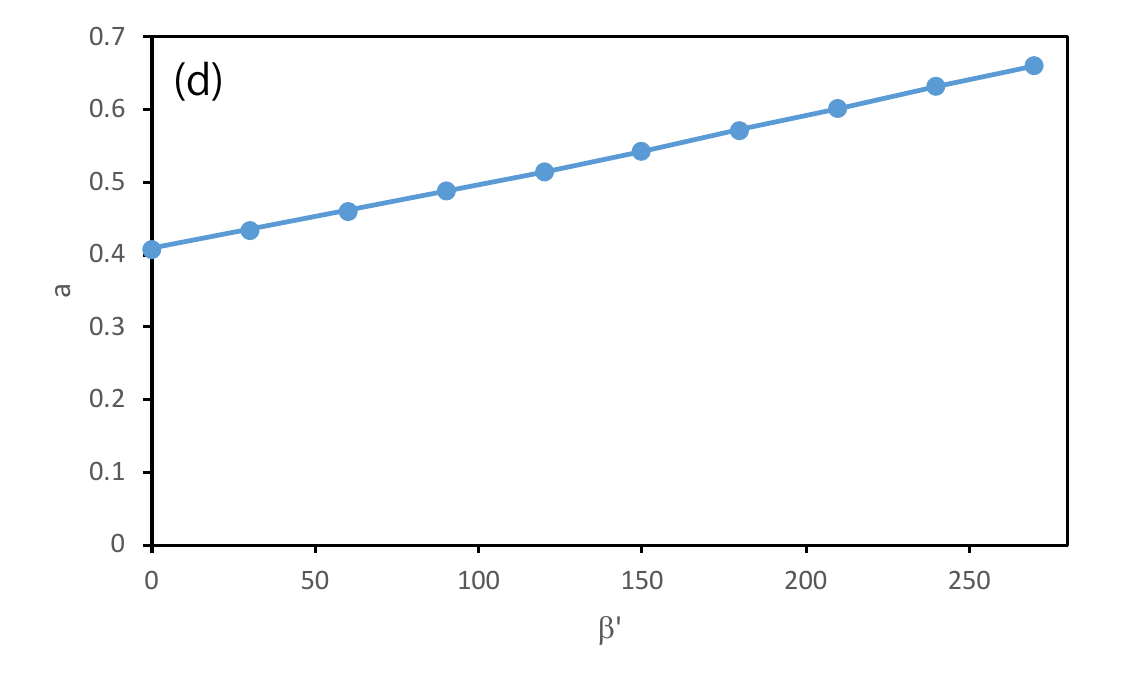}
  \end{center}
 \end{minipage}
\caption{Ridge entropy maximization without considering the degree distribution for the commodity category: FOOD AND LIVE ANIMALS CHIEFLY FOR FOOD. Panels (a) - (d) are objective function $z$, root mean square error, fitting parameter $b$, and fitting parameter $a$, respectively, as a function of control parameter $\beta'$. }
\label{fig:RidgeWOdegree}
\end{figure}

The results of ridge entropy maximization obtained without considering degree distribution are shown in Fig. \ref{fig:RidgeWOdegree}.
Reconstructed bilateral trade $t_{ij}^{(R)}$ was regressed by bilateral trade data $t_{ij}^{(D)}$ using the linear relation $\log t_{ij}^{(R)} = a \log t_{ij}^{(D)} + b$. 
Panels (a) - (d) are objective function $z$, root mean square error, fitting parameter $b$, and fitting parameter $a$, respectively, as a function of control parameter
 $\beta'$.
The root mean square error in panel (b) increases gradually as control parameter $\beta'$ increases contrary to expectation. 
Parameter $b$ in panel (c) and parameter $a$ in panel (d) gradually move toward $0$ and $1$, respectively, as control paramter $\beta'$ increases.

\begin{figure}[htbp]
 \begin{minipage}{0.5\hsize}
  \begin{center}
   \includegraphics[width=60mm]{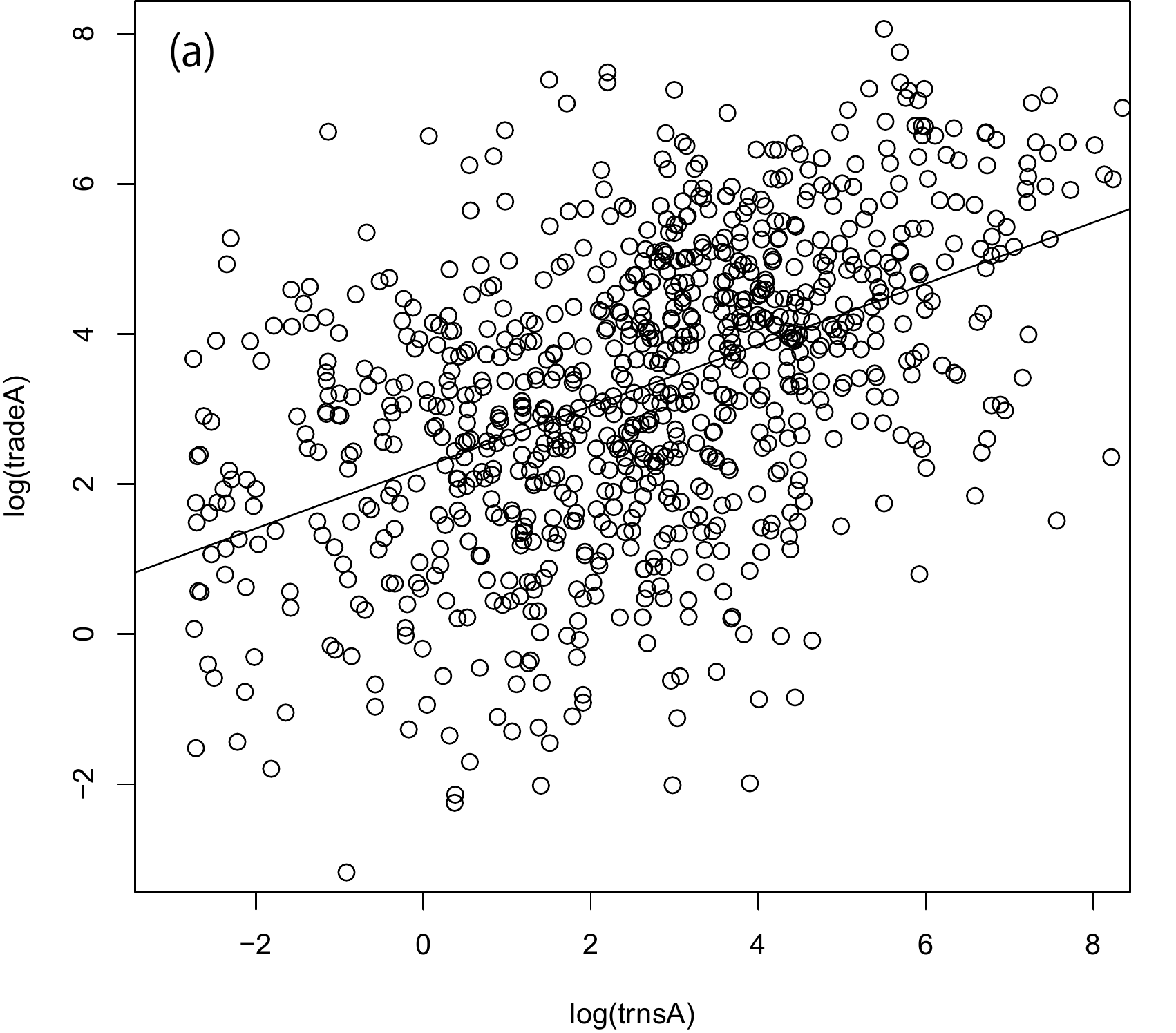}
  \end{center}
 \end{minipage}
 \begin{minipage}{0.5\hsize}
  \begin{center}
   \includegraphics[width=60mm]{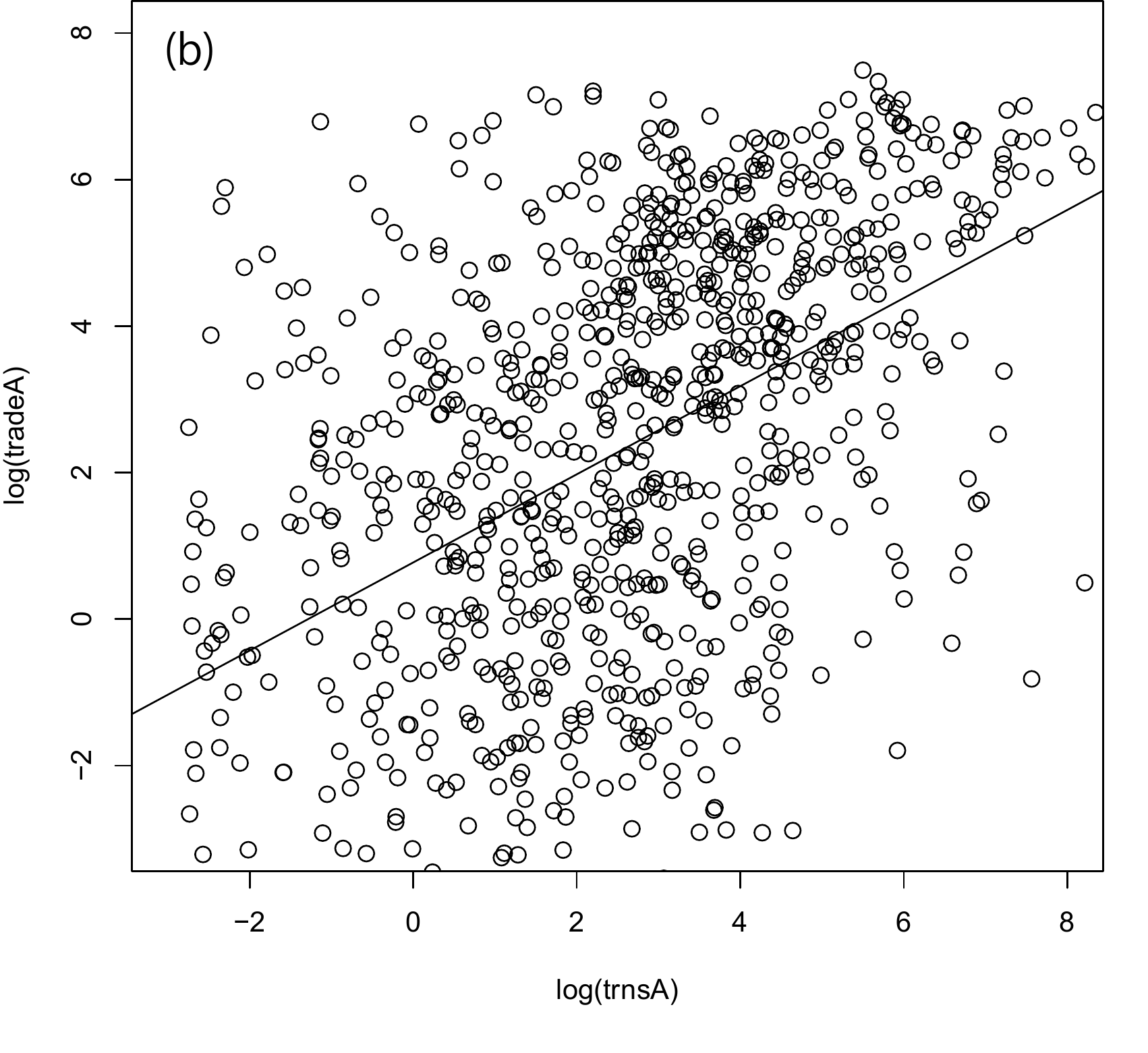}
  \end{center}
 \end{minipage}
\caption{Reconstructed bilateral trade without considering thedegree information for the commodity category: FOOD AND LIVE ANIMALS CHIEFLY FOR FOOD. Panels (a) and (b) are for simple entropy maximization ($\beta'=0$) and ridge entropy maximization ($\beta'=210$), respectively. Fitted line$\log t_{ij}^{(R)} = a \log t_{ij}^{(D)} + b$ is shown. Ridge entropy maximization does not significantly improve linearity.}
  \label{fig:ScatterRidgeWOdegree}
\end{figure}

The realationship between reconstructed trade $t_{ij}^{(R)}$ and trade data $t_{ij}^{(D)}$ are plotted in Fig. \ref{fig:ScatterRidgeWOdegree}.
Panel (a) is for the case of $\beta'=0$ and panel (b) is for the case of $\beta'=210$.

The second type of sparsity, which is characterized as the sparsity of the adjacency matrix, is discussed here.
The accuracy of the reconstruction is expected to be improved using the sparsity of the network. 
The sparsity of the adjacency matrix means that many pairs of nodes are not actually linked. In practice the linked pairs of nodes are searched based on information about in-degree and out-degree distributions using various search algorithms. Among the algorithms, for example, the simplest may be the random search. We assume here that the linked pairs of nodes are obtained at the best result of the search. In the calculation below we eliminated variables $t_{ij}$ for the pairs of nodes with no links in the actual trade data.

\begin{figure}[htbp]
 \begin{minipage}{0.5\hsize}
  \begin{center}
   \includegraphics[width=58mm]{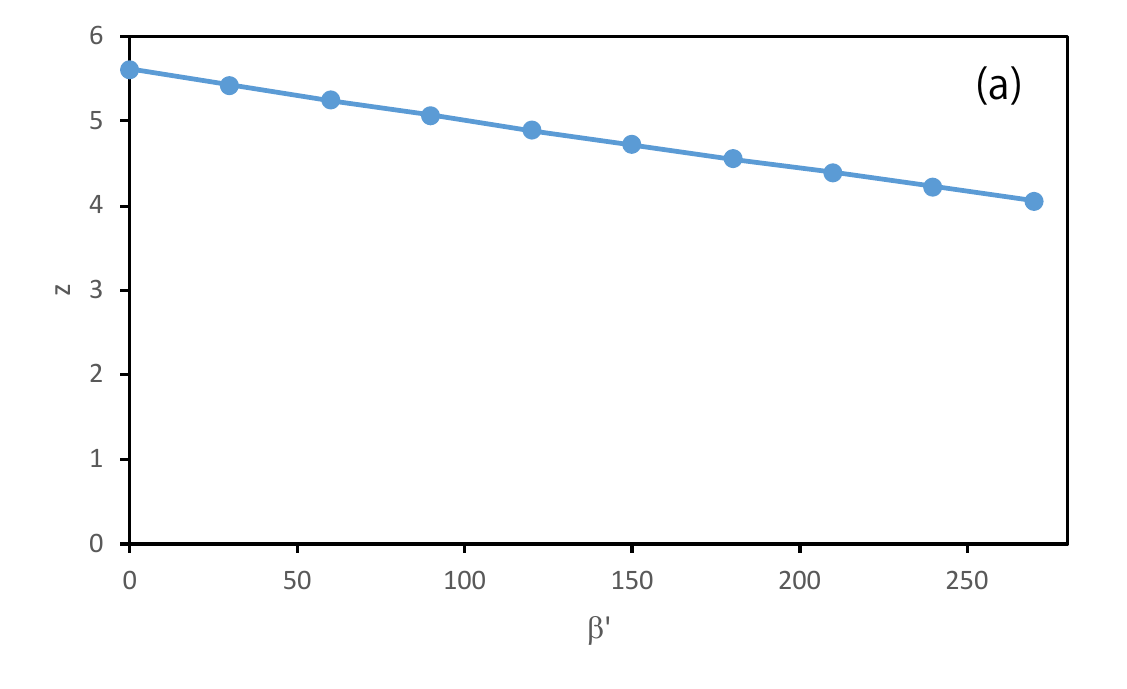}
  \end{center}
 \end{minipage}
 \begin{minipage}{0.5\hsize}
  \begin{center}
   \includegraphics[width=58mm]{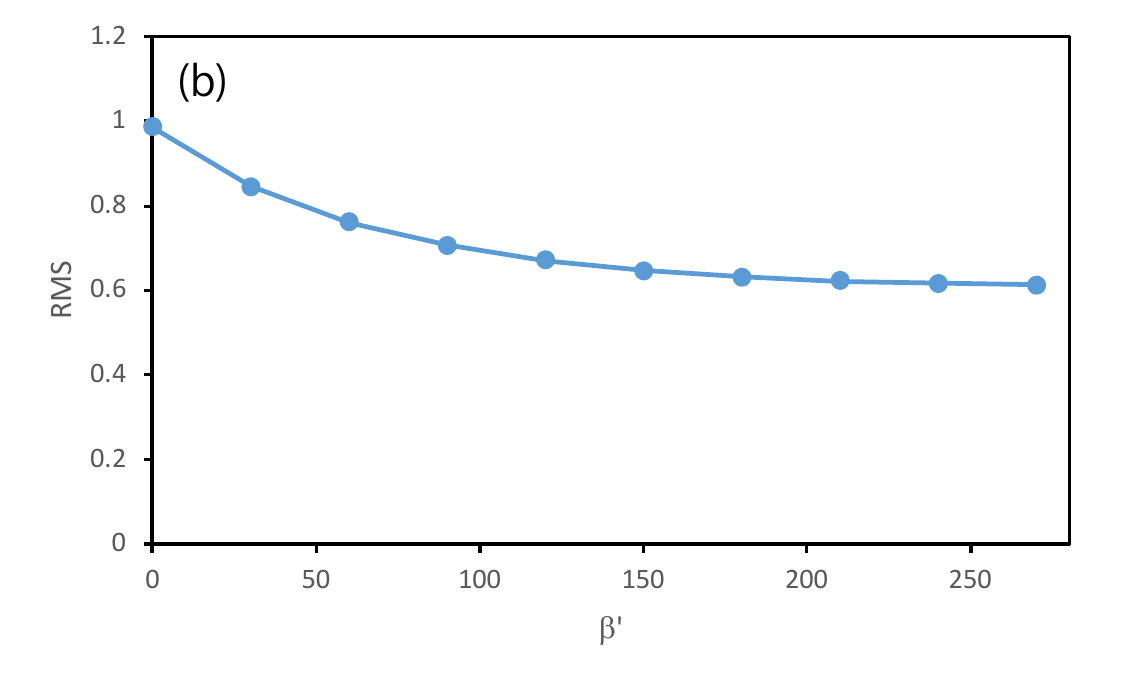}
  \end{center}
 \end{minipage} \\
 \begin{minipage}{0.5\hsize}
  \begin{center}
   \includegraphics[width=58mm]{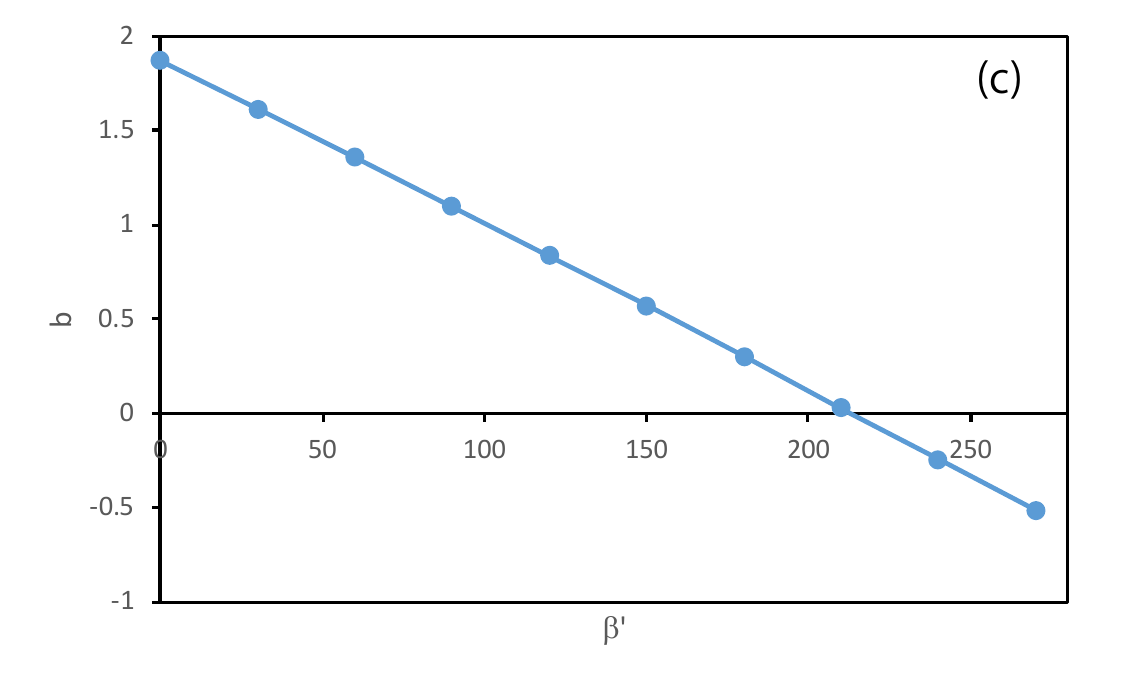}
  \end{center}
 \end{minipage}
 \begin{minipage}{0.5\hsize}
  \begin{center}
   \includegraphics[width=58mm]{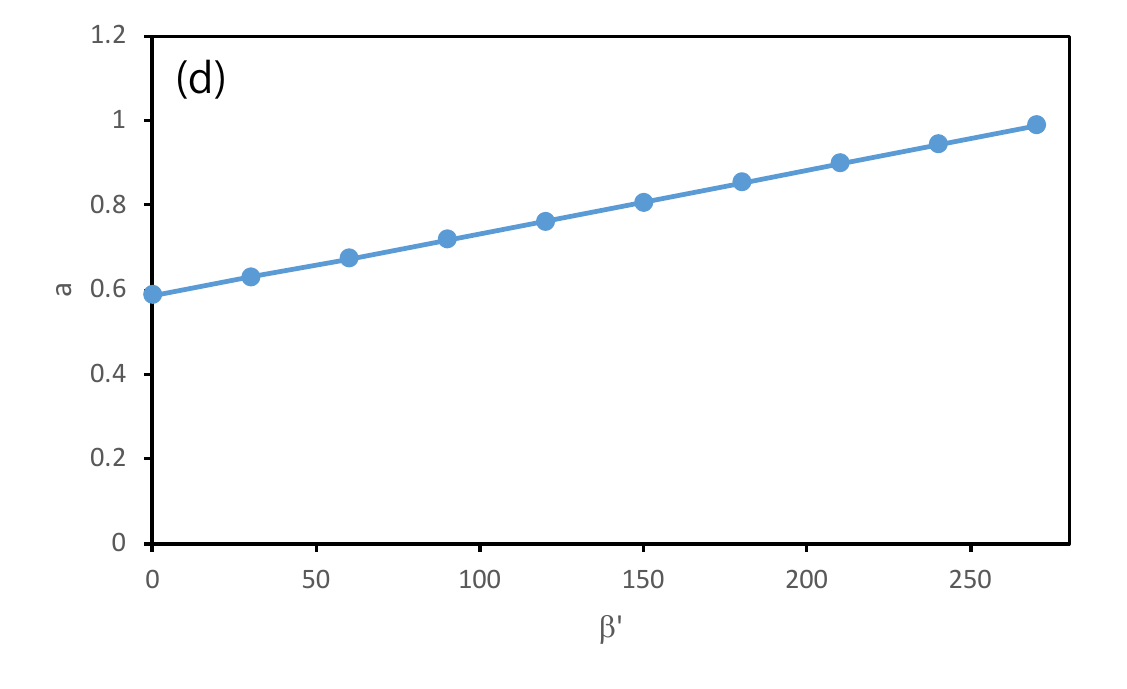}
  \end{center}
 \end{minipage}
\caption{Ridge entropy maximization considering degree distribution for commodity category: FOOD AND LIVE ANIMALS CHIEFLY FOR FOOD. Panels (a) - (d) are objective function $z$, root mean square error, fitting parameter $b$, and fitting parameter $a$, respectively, as a function of control parameter $\beta'$. }
\label{fig:RidgeWdegree}
\end{figure}

The results obtained  from ridge entropy maximization considering the link information are shown in Fig. \ref{fig:RidgeWdegree}.
Reconstructed bilateral trade $t_{ij}^{(R)}$ was regressed by bilateral trade data $t_{ij}^{(D)}$ using the linear relation $\log t_{ij}^{(R)} = a \log t_{ij}^{(D)} + b$. 
Panels (a) - (d) are objective function $z$, root mean square error, fitting parameter $b$, and fitting parameter $a$, respectively, as a function of control parameter
 $\beta'$.
Root mean square error in panel (b) increases gradually as control parameter $\beta'$ decreases according to our expectation. 
Parameter $b$ in panel (c) and parameter $a$ in panel (d) gradually approaches $0$ and $1$, respectively, as control parameter $\beta'$ increases.
These results shows that ridge entropy maximization improves linearity significantly.

The realationship between reconstructed trade $t_{ij}^{(R)}$ and actual trade data $t_{ij}^{(D)}$ are plotted in Fig. \ref{fig:ScatterRidgeWOdegree}.
Panel (a) is for the case of $\beta'=0$ and panel (b) is for the case of $\beta'=210$. 
The RMS error for the case of $\beta'=210$ is equal to $0.620$. 
This shows that considering the link information gives us a better agreement of reconstructed trade to actual trade data compared with the case that does not consider link information.

\begin{figure}[htbp]
 \begin{minipage}{0.5\hsize}
  \begin{center}
   \includegraphics[width=60mm]{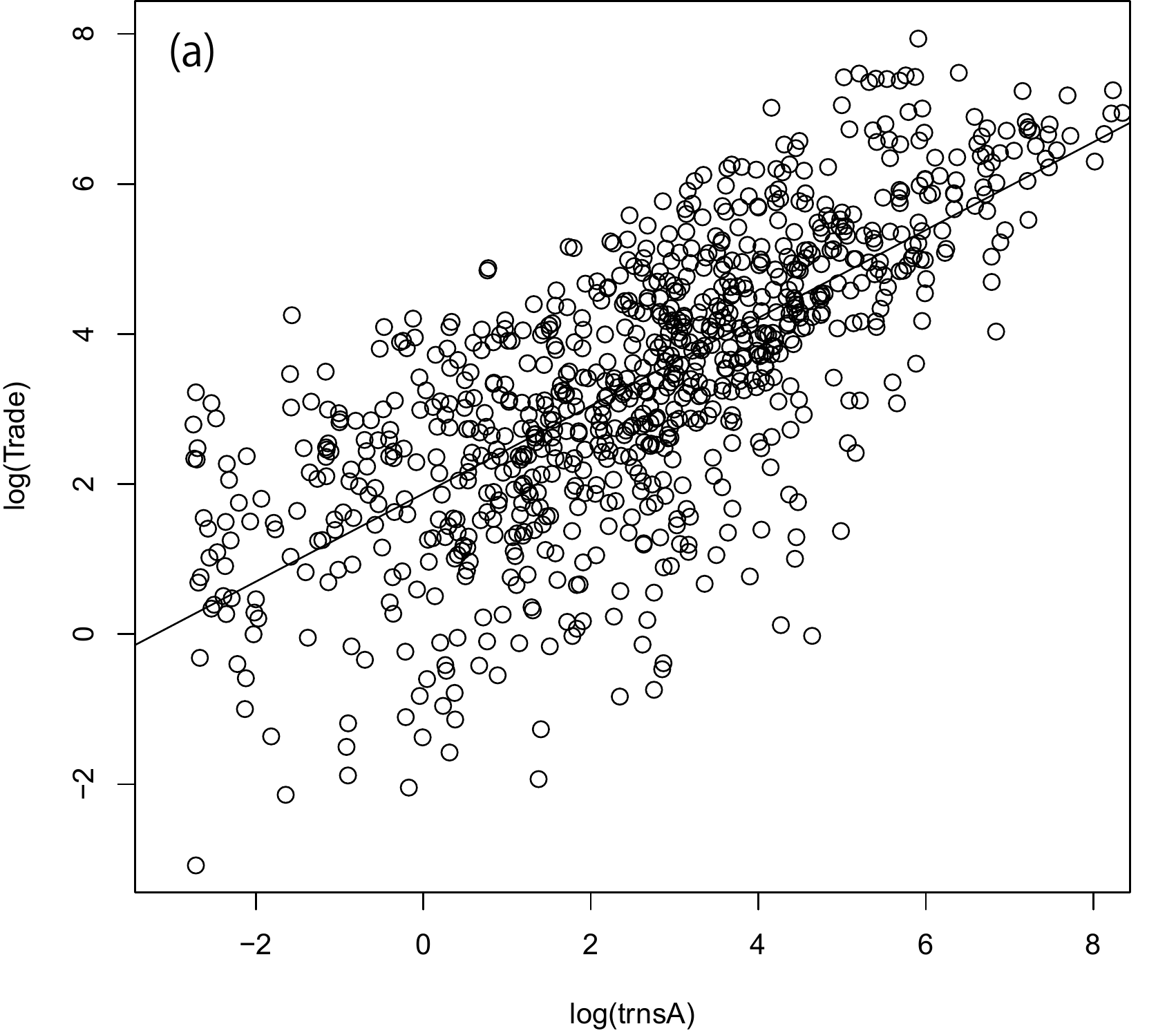}
  \end{center}
  \label{fig:one}
 \end{minipage}
 \begin{minipage}{0.5\hsize}
  \begin{center}
   \includegraphics[width=60mm]{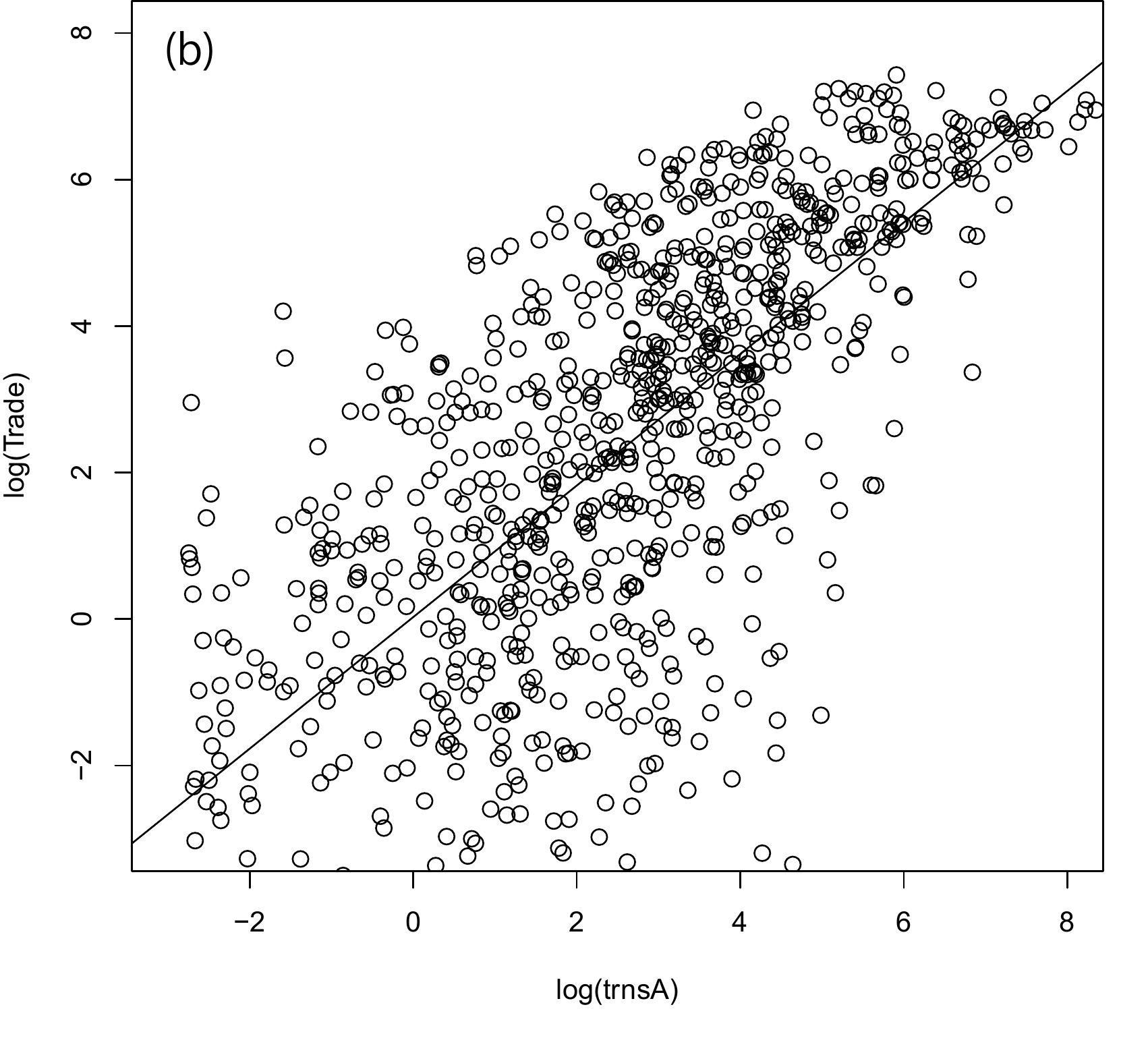}
  \end{center}
  \label{fig:two}
 \end{minipage}
\caption{Reconstructed bilateral trade considering degree information for commodity category: FOOD AND LIVE ANIMALS CHIEFLY FOR FOOD. Panels (a) and (b) are for simple entropy maximization ($\beta'=0$) and ridge entropy maximization ($\beta'=210$), respectively. Fitted line$\log t_{ij}^{(R)} = a \log t_{ij}^{(D)} + b$ is shown. Ridge entropy maximization improves linearity significantly.}
\end{figure}

\subsection{Reconstruction of the Cost Network}
\label{ResultCost}

In this section, we maximize the configuration entropy with the given constraints by estimating the bilateral trade cost which does not appear explicitly in the actual data. The trade cost includes transportation costs, customs duties, and other non-tariff barriers. 

\begin{figure}[htbp]
 \begin{minipage}{0.5\hsize}
  \begin{center}
   \includegraphics[width=60mm]{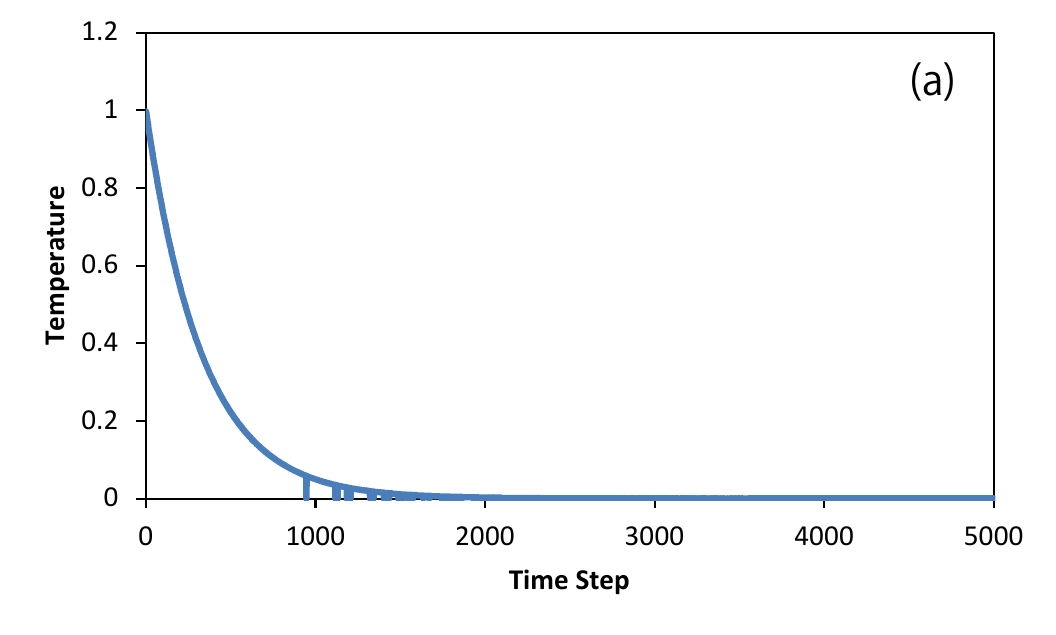}
  \end{center}
  \label{fig:one}
 \end{minipage}
 \begin{minipage}{0.5\hsize}
  \begin{center}
   \includegraphics[width=60mm]{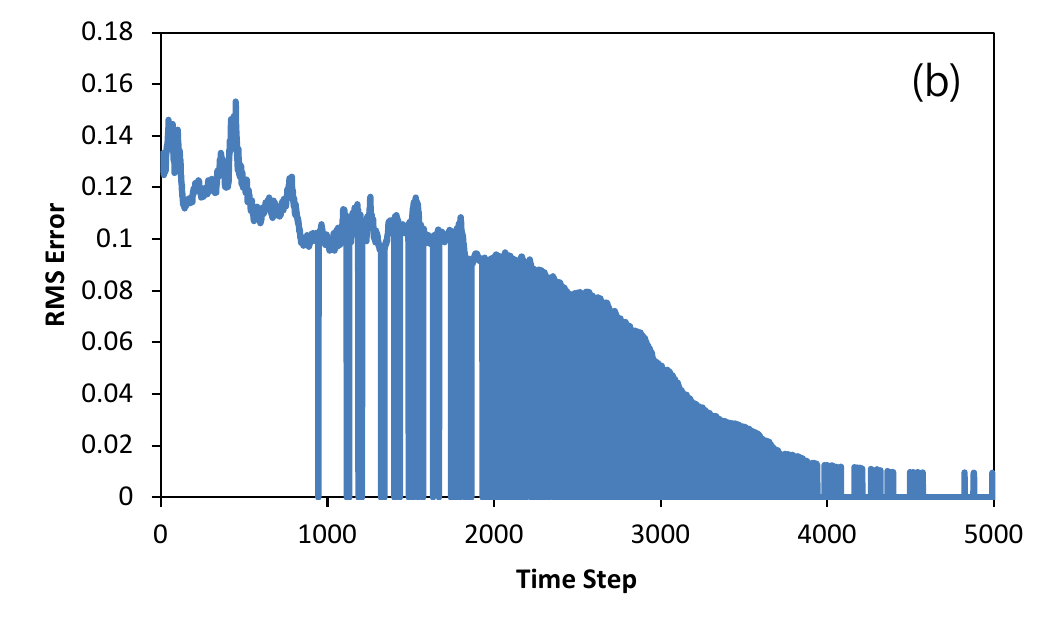}
  \end{center}
  \label{fig:two}
 \end{minipage}
\caption{Temperature and calculated RMS error as a function of the time step of simulated annealing. Panel (a) shows the annealing schedule and panel (b) shows that the root mean square error decreased as the system is annealed. }
\label{fig:TempeatureRMS}
\end{figure}

Reconstruction of the cost network layer was made using simulated annealing according to the algorithm shown in Fig. \ref{Fig:Algorithm}.
Amount of trade $t_{ij}$ is calculated using the iterative relations in Eqs. (\ref{ReconstTrade}) - (\ref{CoeffB}) for the given trade cost $c_{ij}$. 
The annealing schedule is shown in panel (a) of Fig. \ref{fig:TempeatureRMS}. 
Trade cost is initially generated randomly and is changed to a new cost by adding small deviations to all cost pairs. 
At each iteration step, if the root mean square error decreases ($\Delta RMS_n < 0$), the new cost is accepted. But if the root mean square error increases ($\Delta RMS_n > 0$), the probability of the new cost is accepted is low and in the most case it is rejected. 
The root mean square error is shown as a function of the iteration step in panel (b) of Fig. \ref{fig:TempeatureRMS}. 
If the new cost is rejected at iteration step $n$, the RMS error is set equal to zero. Panel (b) shows that the new cost is rejected many times as the system is annealed. 
In spite of the rare acceptance of new costs, RMS was gradually decreased. Costs have converged to the optimal values as the system is annealed.

\subsection{Structural Chnage to the Trade Network}
\label{ResultStrucralChange}

Finally, we simulate structural changes to the international trade network caused by changing the trade tariff reflecting the government's trade policy.  
We selected two commodities. One is category: FOOD AND LIVE ANIMALS CHIEFLY FOR FOOD (abbreviated as FOOD) and the other is category: MACHINERY AND TRANSPORT EQUIPMENT(abbreviated as MACHINERY).
We adopt a trade policy for each category of commodity by reflecting the recent trade protectionist movement away from the establishment of EPAs, such as the TPP. 
The trade policy for commodity category: FOOD is to halve the cost exporting from the U.S. to Japan. The trade policy for commodity category: MACHINERY is to double the cost of exporting from Japan to the U.S..

%
\begin{figure*}
  \begin{center}
  \includegraphics[width=0.7\textwidth]{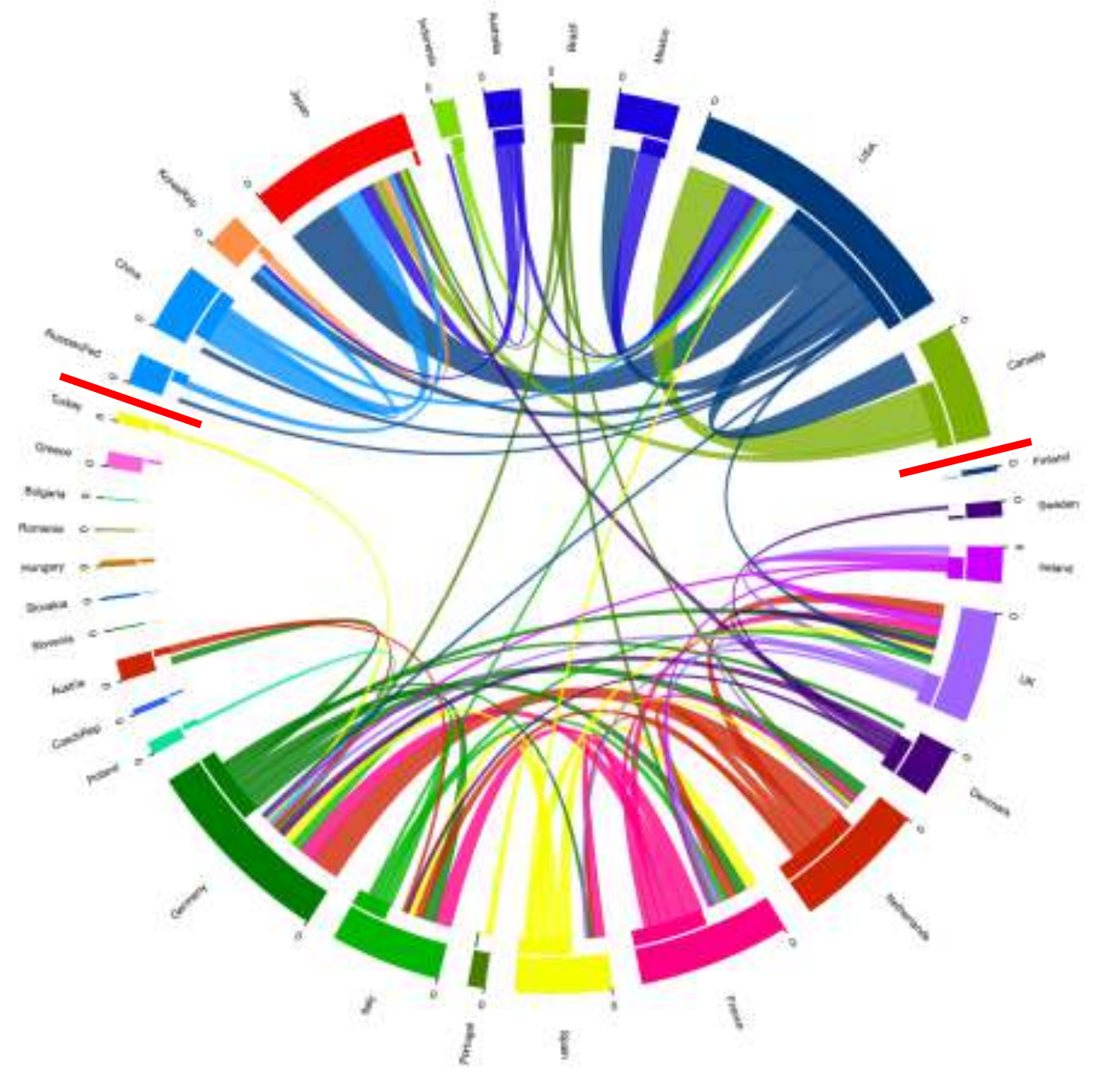}
\caption{Structure of trade network for commodity category: FOOD AND LIVE ANIMALS CHIEFLY FOR FOOD. Two large communities are identified. One corresponds to NAFTA and Asia, and the other corresponds to the EU.}
\label{fig:NetworkFOOD}       
  \end{center}
\end{figure*}

\begin{figure}[htbp]
 \begin{minipage}{0.5\hsize}
  \begin{center}
   \includegraphics[width=60mm]{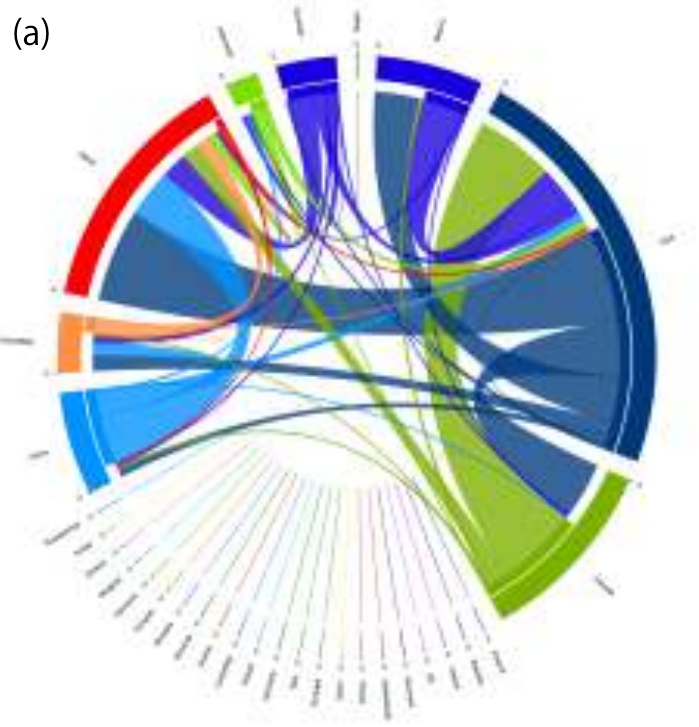}
  \end{center}
 \end{minipage}
 \begin{minipage}{0.5\hsize}
  \begin{center}
   \includegraphics[width=61mm]{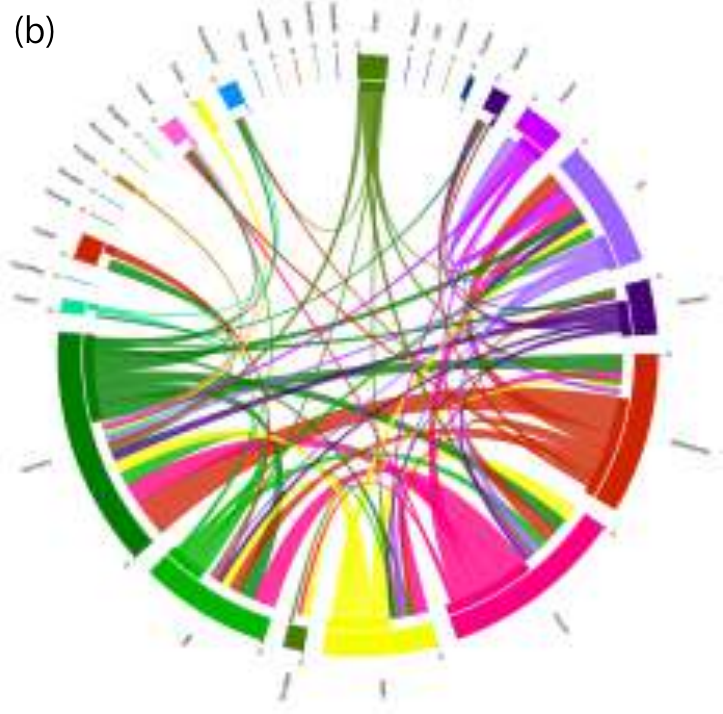}
  \end{center}
 \end{minipage}
\caption{(a) Community 1(NAFTA+Asia) (b) Community 2(EU) }
  \label{fig:NetworkFOODcommunity}
\end{figure}

%
\begin{figure*}
  \includegraphics[width=1.0\textwidth]{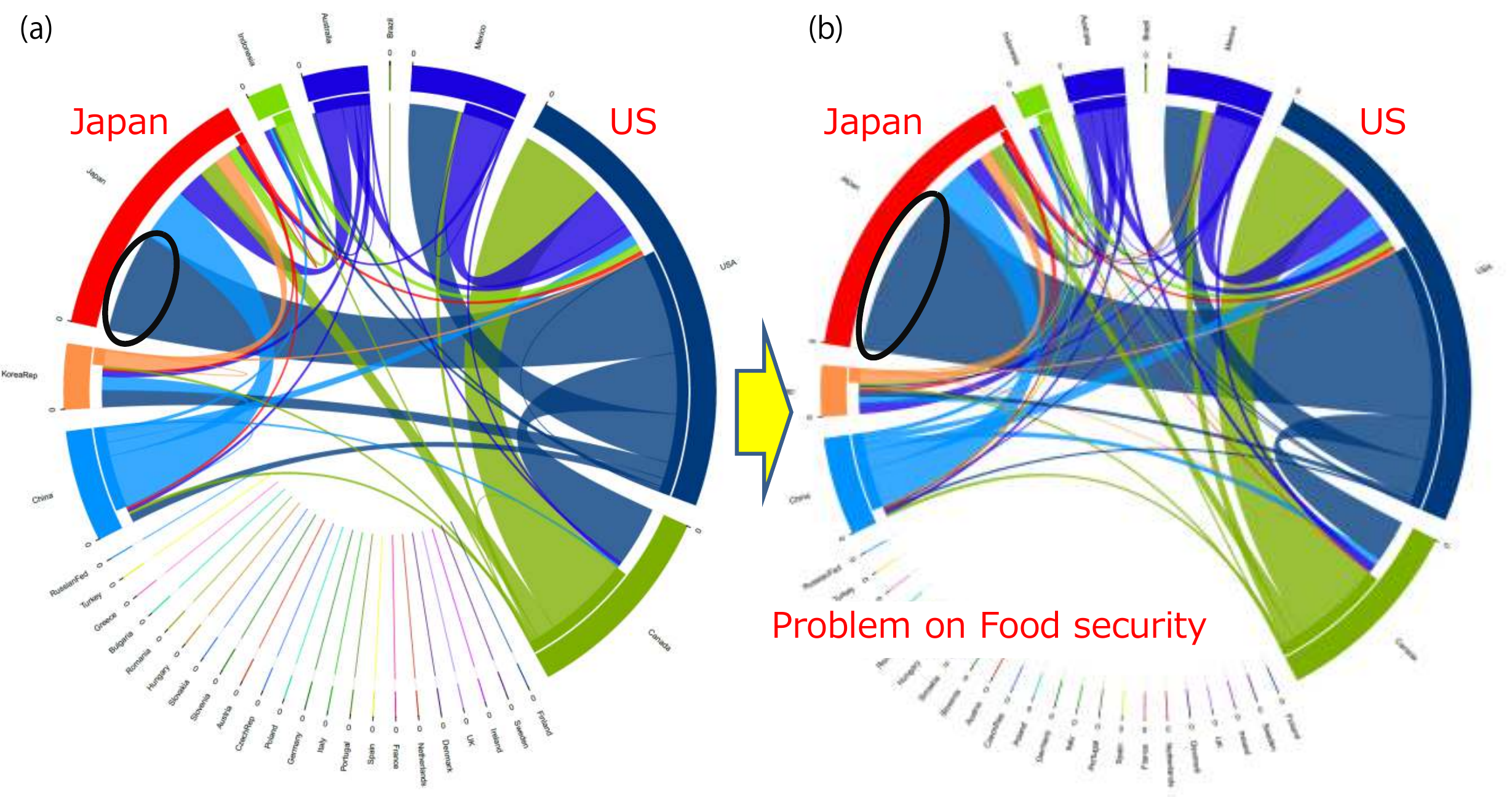}
\caption{Change of Community 1 (NAFTA+Asia) by halving the cost of exporting from the U.S. to Japan for commodity category: FOOD AND LIVE ANIMALS CHIEFLY FOR FOOD. No significant change in community structure is observed, but dependency on the U.S. increases. This might cause problems for Japanese food security.}
\label{fig:NetworkFOODpolicyChange}
\end{figure*}

The structure of trade the network for commodity category: FOODis shown in Fig. \ref{fig:NetworkFOOD}.
Two large communities are clearly identified in Fig. \ref{fig:NetworkFOOD}. the boundaries of the communities are indicated by thick red lines. The two communities are connected by a few thin lines.
Figure \ref {fig:NetworkFOODcommunity} shows that one community corresponds to NAFTA and Asia in panel (a), and the other community corresponds to the EU in panel (b).
Panel (a) shows that Canada and Mexico export food to the U.S. and the U.S. and China export food to Japan. Note that Japan imports almost all of its FOOD.
Panel (b) depicts the interdependence in the EU region. Germany and France export as much FOOD as imports. The Netherlands' food exports exceed its imports. Italy imports more food than it exports. Note that the U.K. imports almost two-thirds of its FOOD.

The change in Community 1 (NAFTA+Asia) by halving the cost of exporting from the U.S. to Japan for commodity category: FOOD is shown in \ref{fig:NetworkFOODpolicyChange}.
Panel (a) depicts trade before the change of policy and panel (b) after halving export costs. FOOD import from the U.S. to Japan is marked by a circle.
The number of communities is not changed, but imports of FOOD  from the U.S. to Japan rise drastically. This might cause problems on Japanese food security.

\begin{figure*}
  \begin{center}
  \includegraphics[width=0.7\textwidth]{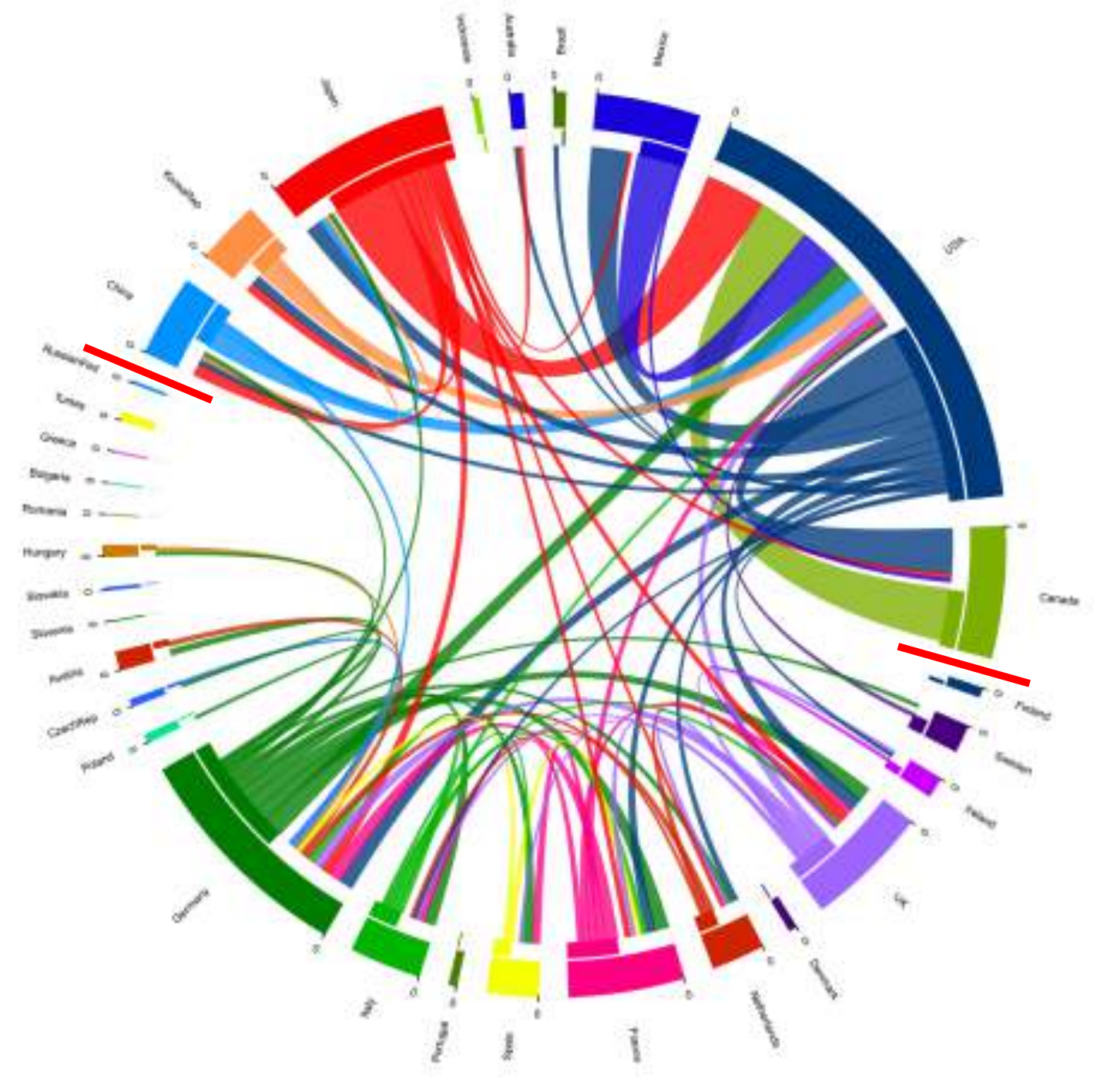}
\caption{Structure of trade network for commodity category: MACHINERY AND TRANSPORT EQUIPMENT. Two large communities are identified again. One corresponds to NAFTA and Asia, and the other corresponds to the EU.}
\label{fig:NetworkMACHINERY}       
  \end{center}
\end{figure*}

\begin{figure}[htbp]
 \begin{minipage}{0.5\hsize}
  \begin{center}
   \includegraphics[width=60mm]{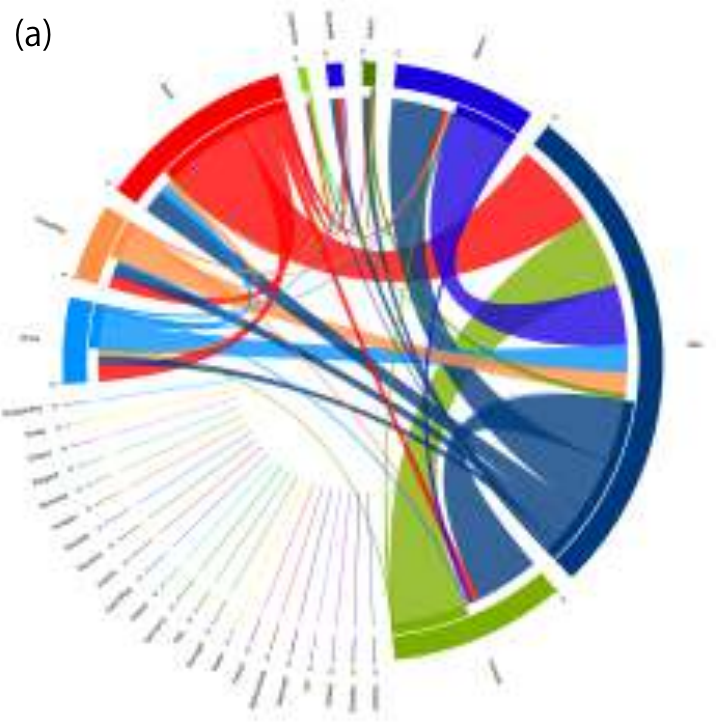}
  \end{center}
 \end{minipage}
 \begin{minipage}{0.5\hsize}
  \begin{center}
   \includegraphics[width=61mm]{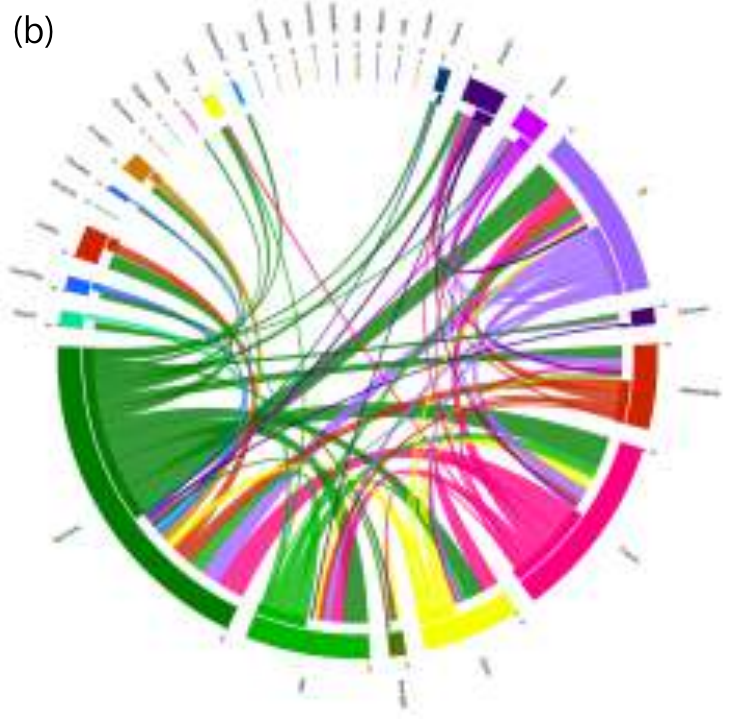}
  \end{center}
 \end{minipage}
\caption{ (a) Community 1 (NAFTA+Asia), (b) Community 2 (EU) }
\label{fig:NetworkMACHINERYcommunity}       
\end{figure}

The structure of the trade network for commodity category: MACHINERY is shown in Fig. \ref{fig:NetworkMACHINERY}. Two large communities are again identified in Fig. \ref{fig:NetworkMACHINERY}. The boundaries of the communities are indicated by thick red lines. The two communities are connected by a few thin lines. However, we note that the lines between Japan and several countries in the EU and the line between Germany and the U.S. are relatively thick.
Figure \ref{fig:NetworkMACHINERYcommunity} shows that one community corresponds to NAFTA and Asia in panel (a), while the other community corresponds to the EU in panel (b).
Panel (a) shows that the U.S. exports MACHINERY to Canada and Mexico and Canada and Mexico export almost same amount to the US. Japan exports large amounts of MACHINERY to the U.S..  
Panel (b) shows the interdependence in the EU region again. Germany, France, the U.K., Italy, and the Netherlands export MACHINERY in  amounts equvalent to their imports. 

\begin{figure*}
  \includegraphics[width=1.0\textwidth]{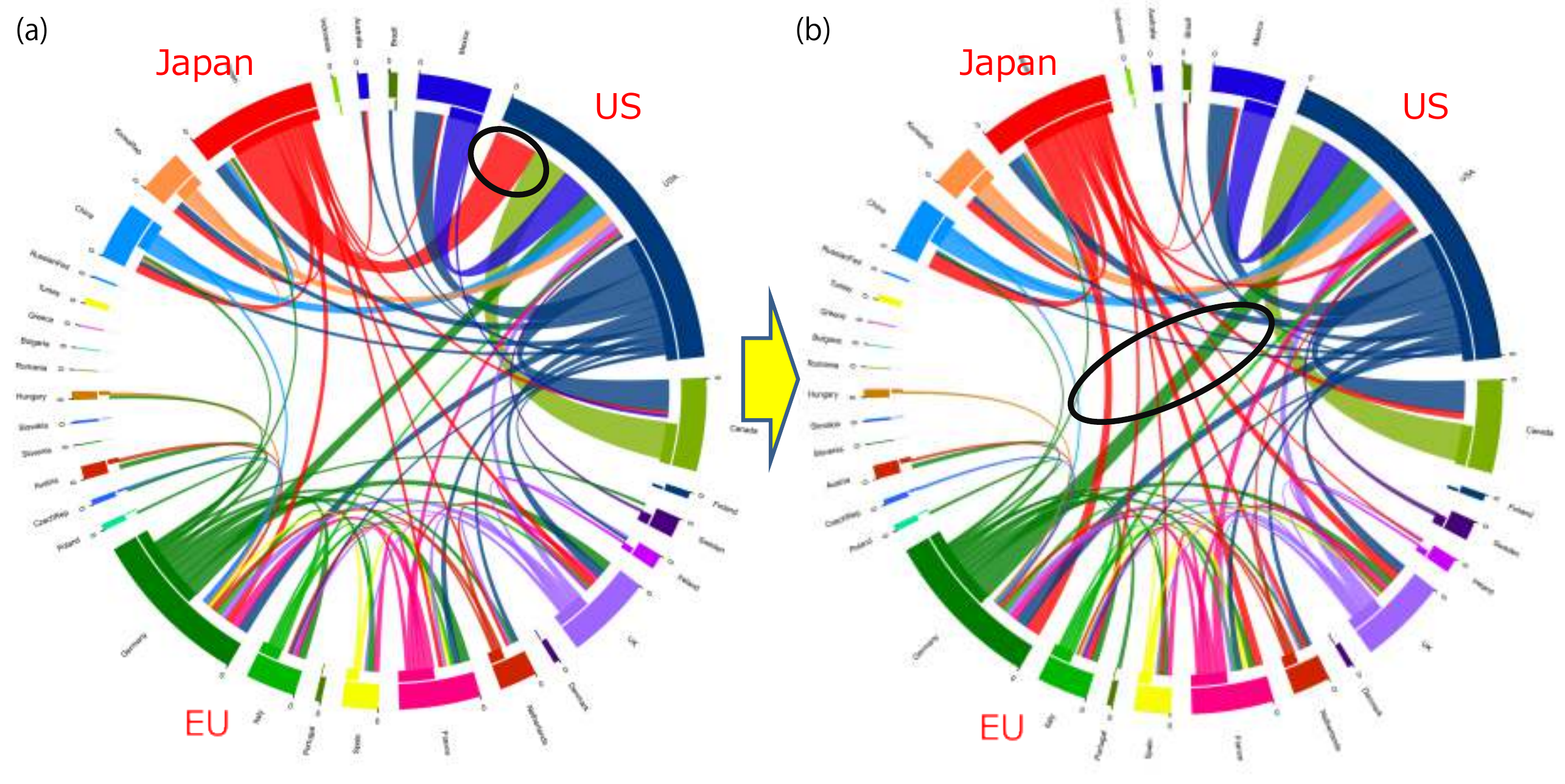}
\caption{Community 1(NAFTA+Asia) and Community 2 (EU) come together as one by doubling the cost of exporting from Japan to the U.S. for commodity category: MACHINERY AND TRANSPORT EQUIPMENT. This might make the EU automotive market more competitive.} 
\label{fig:NetworkMACHINERYpolicyChange}       
\end{figure*}

Figure \ref{fig:NetworkMACHINERYpolicyChange} shows that Community 1(NAFTA+Asia) and Community 2 (EU) come together as one ifthe cost of exporting from Japan to the U.S. are doubled for commodity category: MACHINERY. 
Panel (a) depicts trade before the change in policy and panel (b) shows it after doubling the cost of exporting from Japan to the US. Exports of MACHINERY from Japan to the U.S. are circled.
Panel (b) shows that exports of MACHINERY from Japan to the US decline drastically and exports to countries in the EU region increase.
This might could make the EU automotive market more competitive.

%

%

\section{Conclusion}
\label{Conclusion}

The interdependent nature of the global economy has strongthened with increases in international trade and investment. 
The Japanese economy would achieve higher economic growth under free trade created by the establishment of EPAs such as the TPP. 
We proposed a new model to reconstruct the international trade network and associated cost network by maximizing entropy based on local information about inward and outward trade. 
We showed that a trade network could be reconstructed successfully using the proposed model. 
In addition to these reconstructions, we simulated structural changes in the international trade network caused by changing trade tariffs in the context of a government's trade policy.
We adopted a policy scenario for commodity category: FOOD in which the cost of exporting from the U.S. to Japan was halved. 
The trade policy scenario for commodity category: MACHINERY was to double the cost of exporting from Japan to the U.S.. 
the simulation for the FOOD category shows that imports of FOOD  from the U.S. to Japan rose markedly. This might cause problems for Japanese food security. 
On the other hand, the simulation for the MACHINERY category shows that exports from Japan to the U.S. decreased drastically and exports to countries in the EU region increased. 
This might make the EU automotive market more competitive.

\begin{acknowledgements}
The present study was supported in part by the Ministry of Education, Science, Sports, and Culture, Grants-in-Aid for Scientific Research (B), Grant No. 17904923 (2017-2019) and (C), Grant No. 26350422 (2014-16). This study was also supported by MEXT as Exploratory Challenges on Post-K computer (Studies of Multi-level Spatiotemporal Simulation of Socioeconomic Phenomena). 
\end{acknowledgements}



\end{document}